\documentclass[nofootinbib,aps,preprint,superscriptaddress]{revtex4}
\usepackage{amsmath}
\usepackage{bm}
\usepackage{epsfig}
\newcommand{\nn}{\nonumber} 

\newcommand{\bea}{\begin{eqnarray}}
\newcommand{\eea}{\end{eqnarray}}

\begin{document}

\baselineskip 3.0ex
\vspace*{18pt}
%%%%%%%%%%%%%%%%%%%%%%%%%%%%%%%%%%%%%%%%%%
%Define Title, Author, Address, Preprint#

%\preprint{\vbox{ \hbox{CMU-HEP-03-06}   \hbox{FERMILAB-Pub-03/069-T} }}

\title{Color Octet Scalar Bound States at the LHC}

\author{Chul Kim\footnote{Electronic address: chul@phy.duke.edu}}
\affiliation{Department of Physics, 
	Duke University, Durham,  
	NC 27708\vspace{0.5cm}}

\author{Thomas Mehen\footnote{Electronic address: mehen@phy.duke.edu}}
\affiliation{Department of Physics, 
	Duke University, Durham,  
	NC 27708\vspace{0.5cm}}

%\date{\today}

%%%%%%%%%%%%%%%%%%%%%%%%%%%%%%%%%%%%%%%%%%
%Create the title page

\begin{abstract}\vspace*{18pt}
\baselineskip 3.0ex
One possible extension of the Standard Model scalar sector includes
$SU(2)_L$ doublet scalars that are color octets rather than singlets. We focus on models
in which the couplings to fermions are consistent with the principle
of minimal flavor violation (MFV), in which case these color octet scalars
couple most strongly to the third generation of quarks. 
When the Yukawa coupling of color octet scalars to Standard Model fermions is less than unity, 
these states can live long enough to bind into color-singlet spin-0 hadrons, which we
call octetonia. 
In this paper, we consider the phenomenology of octetonia at the Large Hadron
Collider (LHC). Predictions for their production via gluon-gluon fusion
and their two-body decays 
into Standard Model gauge bosons, Higgs bosons, and $\bar{t}t$ 
are presented.

\end{abstract}

\maketitle

%%%%%%%%%%%%%%%%%%%%%%%%%%%%%%%%%%%%%%%%%%
%\tighten
%\newpage
%%%%%%%%%%%%%%%%%%%%%%%%%%%%%%%%%%%%%%%%%%
%Main body of the paper

A major goal of the physics program at the Large Hadron Collider (LHC) is to  experimentally probe the
mechanism of electroweak symmetry breaking. Although electroweak symmetry breaking can be 
accomplished by the Higgs mechanism with a scalar doublet transforming in the $\bf (1,2)_{1/2}$
representation of the $SU(3)\times SU(2)_L\times U(1)_Y$ gauge group, it is possible that a scalar
sector with a richer structure will be revealed at the LHC. An important constraint on an extended
scalar sector comes from the absence of flavor changing neutral currents.  A natural way to
ensure that theories of new physics are consistent with these  constraints is to invoke the principle
of Minimal Flavor Violation (MFV)~\cite{Chivukula:1987py,D'Ambrosio:2002ex}. MFV  requires that the
Standard Model Yukawa matrices are the sole source of the violation  of the $SU(3)^5$ flavor symmetry
that would be present  in the absence of fermion mass terms. If one invokes MFV as a guiding principle
for models of new physics, possible extensions of the scalar sector are very highly constrained.
Besides the Standard Model Higgs field, only scalars in $\bf (8,2)_{1/2}$
represention can have Yukawa couplings that are consistent with MFV~\cite{Manohar:2006ga}.

Color-octet scalars also arise in $SU(5)$ grand unified theories
if the fundamental theory contains scalars in the $\bf 45$ of $SU(5)$~\cite{Dorsner:2007fy,Perez:2008ry,FileviezPerez:2008ib}.
Higgs fields in this representation are invoked in many $SU(5)$ models to obtain a realistic fermion mass spectrum.
In adjoint $SU(5)$ GUT's, color octet scalars are actually required to be light $(\lesssim 100 \,{\rm TeV})$ in order to obtain
unified couplings~\cite{Perez:2008ry,FileviezPerez:2008ib}. Color octet scalars can also arise in theories
of Pati-Salam unification~\cite{Povarov:2007nh,Popov:2005wz} as well as cosmological models of dark energy \cite{Stojkovic:2007dw}.

All of this motivates serious consideration of the collider signals of color octet scalars.
The LHC phenomenology of color octet scalars has been studied recently in Refs.~\cite{Gresham:2007ri, Gerbush:2007fe}.
In order to be consistent with MFV, the Yukawa couplings of these color octet scalars to Standard Model fermions
  must be~\cite{Manohar:2006ga}
\begin{eqnarray}\label{yukawa}
{\cal L} = - \eta_U \, g_{ij}^U \bar{u}_{R i} T^A Q_{L j} S^A - \eta_D \, g_{ij}^D \bar{d}_{R i} T^A Q_{L j} S^{A\,\dagger} \, ,  
\end{eqnarray}
where 
\bea
S^A =\left( \begin{array}{c} S^{A+} \\ S^{A0}\end{array}\right)
\eea 
are the color octet scalars, $Q_{L j}$ is the Standard Model $SU(2)_L$ doublet of quarks,
$u_R$ and $d_R$ are the right-handed Standard Model quark fields, $g^{U,D}_{ij} =\sqrt{2}\,M^{U,D}_{ij}/v$,
where $M^{U,D}_{ij}$ are the Standard Model quark mass matrices and $v/\sqrt{2}$ is the vacuum expectation value of the Higgs doublet.
In addition to the Yukawa couplings listed above the scalar octets have gauge couplings to Standard 
Model gauge bosons, and there is a scalar potential which allows for scalar octet self-interactions
as well as scalar octet couplings to  the Standard Model Higgs~\cite{Manohar:2006ga}. In general, there
are three mass eigenstates, $S^{A\pm}$, $S^{A0}_R = \sqrt{2} \, {\rm Re}\,S^{A0}$, and $S^{A0}_I = \sqrt{2}\, {\rm
Im}\,S^{A0}$, with masses given by
\bea\label{masses}
m^2_{S^\pm} &=& m_S^2+\lambda_1 \frac{v^2}{4} \nn \\
m^2_{S_R^0} &=& m_S^2+(\lambda_1 +\lambda_2 +2 \lambda_3)\frac{v^2}{4} \nn \\
m^2_{S_I^0} &=& m_S^2+(\lambda_1 +\lambda_2 -2 \lambda_3)\frac{v^2}{4} \, ,
\eea
where $m_S$, $\lambda_1$, $\lambda_2$, and $\lambda_3$ are parameters appearing in the scalar potential. 
If custodial $SU(2)$ symmetry is imposed on the potential, $\lambda_2=2\lambda_3$, and the $S^{A\pm}$ and $S^{A0}_I$ are degenerate. 

If $\eta_U$ and $\eta_D$ are both order unity, the dominant decays 
of the $S^A$ are $S^{A+} \to t \bar{b}$ and $S^{A0}_{R,I} \to \bar{t} t$, or $S^{A0}_{R,I} \to \bar{b} b$, if the $S^{A0}_{R,I}$ mass lies below
$2 m_t$.  In this case, to search for color octet scalars, one
would look for modifications to Standard Model predictions for final states with multiple heavy flavor jets.
Ref.~\cite{Gerbush:2007fe} obtained  a lower bound of $\sim 200$ GeV for the color octet masses by 
studying the process $gg\to S^A S^A \to b\bar{b} b\bar{b}$, which is constrained by the 
CDF search for $H\to \bar{b}b$ plus associated $b$ jets in Ref.~\cite{CDF8954}.

In this paper we investigate the possibility of producing color singlet  bound states of these color octet scalars,
which we will call octetonia.
For this to occur, the scalar octet must live long enough so that it has time
to form a bound state prior to decaying. For example, we have observed top quarks in hadron colliders, but 
not toponium resonances because the top quark decays too quickly.  If they exist, octetonia are heavy enough that they can be 
reasonably modelled as quasi-Coulombic bound states, therefore the binding energy of the ground state octetonium
is expected to be $\approx m_S N_c^2 \alpha_s^2(m_S v)/4$, where $m_S$ is the color octet scalar mass, $N_c$ is the number of colors, 
and $v$ is the typical velocity of the scalar in the bound state.~\footnote{The typical velocity of the constituents of a quasi-Coulombic bound state of octets in QCD is estimated by demanding
$v = N_c \alpha_s(m_S v)$~\cite{Luke:1999kz}, which, for the one-loop expression for $\alpha_s$,  is solved by 
\bea
v= \frac{2 N_C \pi}{\beta_0 w[2 N_c \pi m_S/(\beta_0 \Lambda_{\rm QCD})] }\, ,
\eea
where $w[x]$ is the Lambert W (or Product Log) function. For $m_S = 200 - 1200$ GeV, we find $v=0.37 - 0.30$
and $\alpha_S(m_S v) = 0.12-0.10$.}
The factor $N_c$ appears here because it is 
the color factor in the Coulomb potential for two particles in the $\bf 8$ representation.
If $\tau_{\rm form}$ is the time scale for formation 
of the bound state (after the $S^A$ pair has been produced), and $\tau_{\rm life}$ is the lifetime of the two particle state, then, 
using  the decay widths for $S^{A+} \to t \bar{b}$ calculated in Ref.~\cite{Manohar:2006ga}
we find for bound states of $S^{A\pm}$,
 \bea\label{formlife}
 \frac{\tau_{\rm form}}{\tau_{\rm life}} = \frac{2\Gamma[S^{A+}]}{B.E.} \sim 0.08\, |\eta_U|^2-0.6 \,|\eta_U|^2 \, ,
\eea
for $m_{S^\pm}$ in the range $200-500 \, {\rm GeV}$. Here $B.E.$ is the binding energy. For comparison, the corresponding ratio for top quarks
yields $\tau_{\rm form}/\tau_{\rm life} \approx 3$.
For bound states containing neutral scalars, $\tau_{\rm form}/\tau_{\rm life}$ depends strongly on $m_{S^0_R}$ and 
$m_{S^0_I}$. If both masses are below the $t\bar{t}$ threshold, $342\, {\rm GeV}$, the $S^{A0}_{R,I}$ must decay to $b \bar{b}$ and then 
$\Gamma[S^{A0}]$ is suppressed by $(m_b/m_t)^2$ and $\tau_{\rm form}/\tau_{\rm life} \sim 6 \times 10^{-4} |\eta_D|^2$. In this scenario
bound states clearly can form if $\eta_D$ is of order unity. For  $m_{S^0_{R,I}}$ above 400 GeV, the estimate
for $\tau_{\rm form}/\tau_{\rm life}$ is similar to Eq.~(\ref{formlife}).
In any case, we see that when $\eta_U < 1$ the criterion for bound state formation is satisfied.

Ref.~\cite{Gresham:2007ri} studied the impact  of one loop diagrams with virtual color octet scalars on the Standard Model 
prediction for $R_b= \Gamma[Z^0 \to b\bar{b}]/\Gamma[Z^0 \to \,{\rm hadrons}]$.  Their results give an upper bound 
 on $\eta_U$ as a function of $m_{S^\pm}$. For $m_{S^\pm} = 200$ GeV the 1$\sigma$ (2$\sigma$)
upper bound is $|\eta_U| < 0.34$ ($|\eta_U| < 0.79$), while for $m_{S^\pm} = 500$ GeV the upper bound
is $|\eta_U| < 0.62$ ($|\eta_U| < 1.45$), so for color octet scalars whose masses lie just beyond the existing
upper bounds, the Yukawa coupling $\eta_U$ will be small enough to accommodate the production of octetonia.

Like the $S^A$ themselves the octetonia  will decay into heavy quarks. However, the bound states
can also decay via annihilation, much like conventional quarkonium states. The decays we focus on here are the decays 
to two gauge bosons: $gg$, $\gamma \gamma$, $W^+ W^-$, $\gamma Z^0$, and $Z^0 Z^0$. Since the couplings of $S^A$ to gauge
bosons are fixed entirely by gauge invariance, calculation of these decay rates is model independent up to a universal 
factor, $|\psi(0)|^2$, the wave function at the origin squared. This sets the normalization of decay rates and cross sections,
and can be crudely estimated in the quasi-Coulombic approximation. Thus the production cross section (as a function of the
octetonium mass) can be estimated to within a factor of order unity. Furthermore, the relative branching fractions for the 
two-body decays to Standard Model gauge bosons can be reliably calculated. We also calculate the octetonia decays to Higgs boson pairs 
and to $t\bar{t}$. These decays are sensitive to the Yukawa couplings of the color octet scalars as well as parameters
appearing in the color octet scalar potential.

In this paper we will focus on the production and decay of the three octetonia which can be produced directly
via gluon-gluon fusion. These are $O^0_+$,  $O^0_R$, and $O^0_I$, which are color-singlet bound states 
of $S^{A+} S^{A-}$, $S^{A0}_R S^{A0}_R$, and  $S^{A0}_I S^{A0}_I$, respectively. Many other bound states are
possible, for example,  $O^+_R$ which is a boundstate of $S^{A+} S^{A0}_R$ or $O^0_{RI}$ which is a bound state
of $S^{A0}_R S^{A0}_I$. These octetonia
can be  produced in association with 
electroweak gauge bosons, for example, $gg\to O^+_R W^-$ or $gg\to O^0_{RI} Z^0$. These
states cannot decay into to two gluons which is the dominant annihilation decay mode for $O^0_+$, $O^0_R$, and $O^0_I$. Therefore,
octetonia such as $O^+_R$ or $O^0_{RI}$ may appear as two-body resonances in final states with three
electroweak bosons which could make for interesting signals at the LHC. We leave the study of the production and decay of these octetonia
to future work.

For the $O^0_+$,  $O^0_R$, and $O^0_I$, we consider only the production of the lowest radial excitation
of the $S$-wave color singlet state. The wave function at the origin squared for a state with principal quantum number
$n$ will scale as $1/n^3$ so production cross sections for radially excited states will be correspondingly suppressed.
This is also true for partial waves other than $S$-waves. From our calculation of $pp \to O^0_+ \to \gamma \gamma$,
we expect that octetonia will form narrow resonances that will exceed Standard Model background if $m_S$ lies in the range 
200 - 500 GeV, corresponding to octetonia with masses from 400 - 1000 GeV, so we focus on this mass range 
in this paper. Nonobservation of resonances in $\gamma \gamma$, $W^+ W^-$, $Z^0 Z^0$, etc., in this mass range
will significantly increase existing lower bounds on the color octet scalar masses. In our calculations 
 we will take $m_O = 2m_S$, neglecting the binding energies which are expected to be of about 6 (for $m_O=400$ GeV)
 to 16 GeV (for $m_O = 600$ GeV)
in the quasi-Coulombic approximation. (Calculations of superheavy color-octet quarkonium binding energies in a realistic
potential model yield similar estimates~\cite{Hagiwara:1990sq}.) This $\sim 1.5 \%$ shift in the mass of the octetonia
will not significantly affect our predictions for production cross sections or decay rates.

The amplitude for the decay $O^+ \to G G^\prime$, where $G$ and $G^\prime$ are Standard Model gauge bosons,
is obtained by convolving the amplitude for $S^{A+} S^{B-}\to G G^\prime$ with the  wavefunction of the octetonium state, which is
\bea\label{wfn}
|O^0_+(P)\rangle &=& \frac{\delta^{AB}}{\sqrt{N_c^2-1}}\sqrt{\frac{2}{m_O}}\int \frac{d^3 k}{(2\pi)^3} \, \tilde \psi(k)\, |S^{A+}(P/2+k) S^{B-}(P/2-k)\rangle \, ,
\eea
for $O^0_+$. A similar definition holds for $O^0_R$ and $O^0_I$. Here $P^\mu$ is the four-momentum of  $O_+^0$,
$k^\mu$ is the relative momentum of the $S^A$ within the bound state, and $\tilde \psi(k)$ is the momentum space
wavefunction. The prescription for computing the 
production or decay is analogous to standard calculations of quarkonium production or decay. For decays of $O^0_+$, one computes
the amplitude for $S^{A+}(P/2+k) S^{B-}(P/2-k) \to X$, where $X$ is the final state, and then convolves this amplitude 
with the wavefunction in Eq.~(\ref{wfn}). Since $P^0\sim{\cal O}(m_O)$ but $k^\mu \sim{\cal O}(m_S v)$, 
the amplitude is expanded in powers of $k^\mu$. Keeping only the leading term in this expansion leads to the result that 
${\cal M}[O^0_+(P) \to X]$ is equal to the amplitude ${\cal M}[S^{A+}(P/2)S^{A-}(P/2)\to X]$ times $\sqrt{2/m_O} \,\psi(0)$ 
 times a color factor. 

For the  parton level cross section via gluon gluon fusion we obtain 
\bea
\hat{\sigma}[g g \to O^0_+] &=& \frac{9\,\pi^3 \alpha_s^2(2 m_S)}{2 m_S \hat{s}}  |\psi(0)|^2 \, \delta(\hat{s} - m_O^2) \, \nn \\
\hat{\sigma}[g g \to O^0_{\rm R}]  &=&\hat{\sigma}[g g \to O^0_{\rm I}] = \frac{1}{2} \hat{\sigma}[g g \to O^0_+] \, ,
\eea
where $\hat s$ is partonic center of mass energy squared,
and $\alpha_s$ is evaluated at the scale $2 m_S$.
These expressions must be convolved with parton distribution functions to obtain the total cross sections.
To estimate $|\psi(0)|^2$ we use a Coulombic wavefunction with $\alpha_s$ evaluated at the scale $m_s v$,
\bea
|\psi(0)|^2 = \frac{N_c^3 \alpha_s^3(m_S v) m_S^3}{8\pi} \, .
\eea
This is at best a rough approximation to the actual result. Bound states of superheavy quarks were
studied in Ref.~\cite{Hagiwara:1990sq} using a realistic potential model. These authors  showed 
that the quasi-Coulombic approximation is not very 
accurate even for superheavy quarkonium. For example, for quarks with masses of 500 ${\rm GeV}$, taking 
$\Lambda_{\rm QCD} = 200$~MeV, the authors of Ref.~\cite{Hagiwara:1990sq} find 
$1/\langle r(1S)\rangle = 25 \, {\rm GeV}$ for the inverse size of the 
$1S$ quarkonium state in their potential model, while the Coulombic approximation
yields $1/\langle r(1S)\rangle = 35 \, {\rm GeV}$. Since $|\psi(0)|^2$ should scale as
$\langle r(1S)\rangle^{-3}$, this implies $|\psi(0)|^2$ could be smaller than the quasi-Coulombic estimate
by a factor of 2. On the other hand, Ref.~\cite{Hagiwara:1990sq} also showed that $|\psi(0)|^2$ could be significantly affected by Higgs exchange, 
which results in an attractive force between superheavy quarks that
can increase $|\psi(0)|^2$ by a similar factor. Similar considerations hold for color octet
scalars, and there is additional uncertainty because the strength of the color-octet
couplings to the Higgs is unknown, since the $\lambda_i$ in Eq.~(\ref{masses}) cannot 
be determined from the octet scalar masses. Furthermore, there are contact interactions
from the  color-octet scalar potential with unknown coefficients. We expect the quasi-Coulombic estimate
to be correct up to factors of order unity.

\begin{figure}[!t]
\vspace{-1 cm}
\epsfig{file=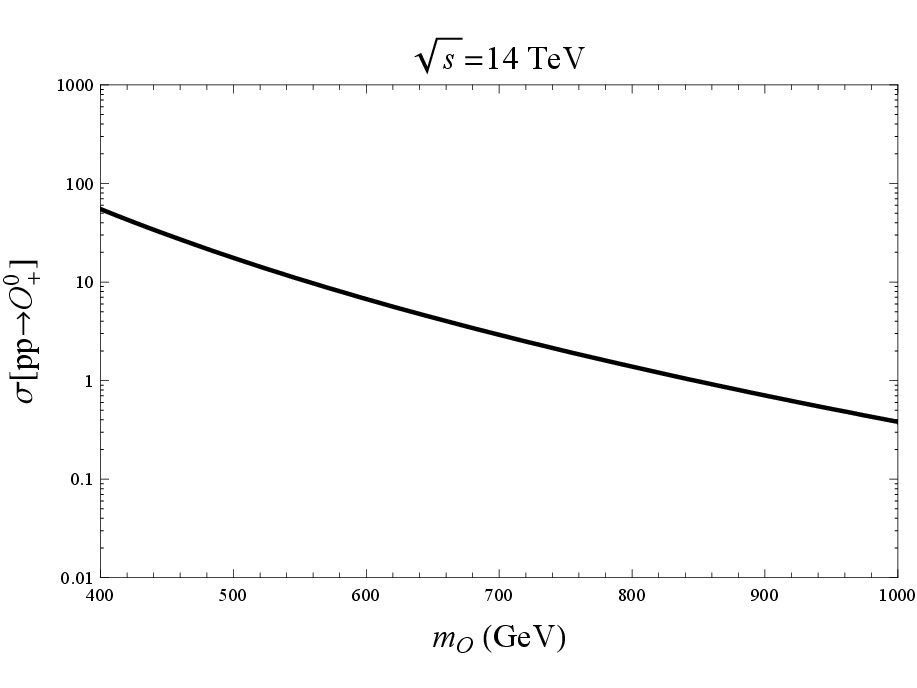, width=16cm}
\vspace{-0.5 cm}
\caption{Total cross section for $pp\to O_+^0  \to X$ at the LHC.  \label{TotalCrossSections}}
\end{figure}

The total cross section for $O^0_+$ production at the LHC $(\sqrt{s} =14$ TeV), is shown in
 Fig.~\ref{TotalCrossSections}. We have used the CTEQ5 parton distribution function~\cite{Lai:1999wy}.
 The production cross section ranges from $\sim 50$ pb for $m_{O_+^0} = 400$ GeV, 
to 1 pb for $m_{O_+^0} = 1000$ GeV. Cross sections for $O^0_R$ and $O^0_I$ are one half as large.
With an integrated luminosity of $\sim 100$ fb$^{-1}$ at the LHC, significant
numbers of octetonia should be produced if they exist in this mass range. 

Next we discuss the two-body decays of octetonia. Unlike the $S^A$ particles,
which must decay to Standard Model fermions, the octetonia can decay to
Standard Model electroweak gauge bosons. $O^0_+$  decays to $W^+W^-, Z^0 Z^0,
\gamma \gamma$, and $\gamma Z^0$, and $O^0_{R,I}$ can decay to $W^+W^-$ and
$Z^0 Z^0$. These channels may be promising for searching for the octetonia
since QCD backgrounds are under better control  than for final states with four
heavy jets. The  octetonia can also decay to $\bar{t}t$ and $gg$ and 
therefore should appear as  resonances in dijet and top quark pair production
cross sections.

CDF and D0 have done searches for resonances in dijets~\cite{Abe:1997hm,Giordani:2003ib}
 and $\bar{t}t$~\cite{CDF,DO} which can be used to rule out 
new physics coupling to these final states.  The upper bounds on
$\sigma[p\bar{p}\to X] \,{\rm Br}[X\to t\bar{t}]$ range from about 2 pb for $m_X =400$ GeV to about 0.5 pb
for $m_X =  900$ GeV. The total cross section for octetonia at the Tevatron is orders of magnitude smaller than these bounds.
For example, at $\sqrt{s} = 1.96$ TeV, we estimate $\sigma [p\bar{p} \to O_+] = 0.062$ pb for $m_{O^0_+} = 400$ GeV and $\sigma [p\bar{p} \to O_+] = 6.5 \times 10^{-6}$ pb at $m_{O^0_+}= 900$ GeV. The contribution
to dijet cross sections from octetonia is also negligible the Tevatron.
The production of octetonia at the Tevatron is suppressed because it is initiated by gluons and because the 
cross section is proportional to $\alpha_s^2 (2 m_S)  \alpha_s^3 (m_S v)$. The gluon rich environment at the LHC
will be a much more promising place to search for these states.

\begin{figure*}[!t]
\vspace{-1cm}
\includegraphics[width=16 cm]{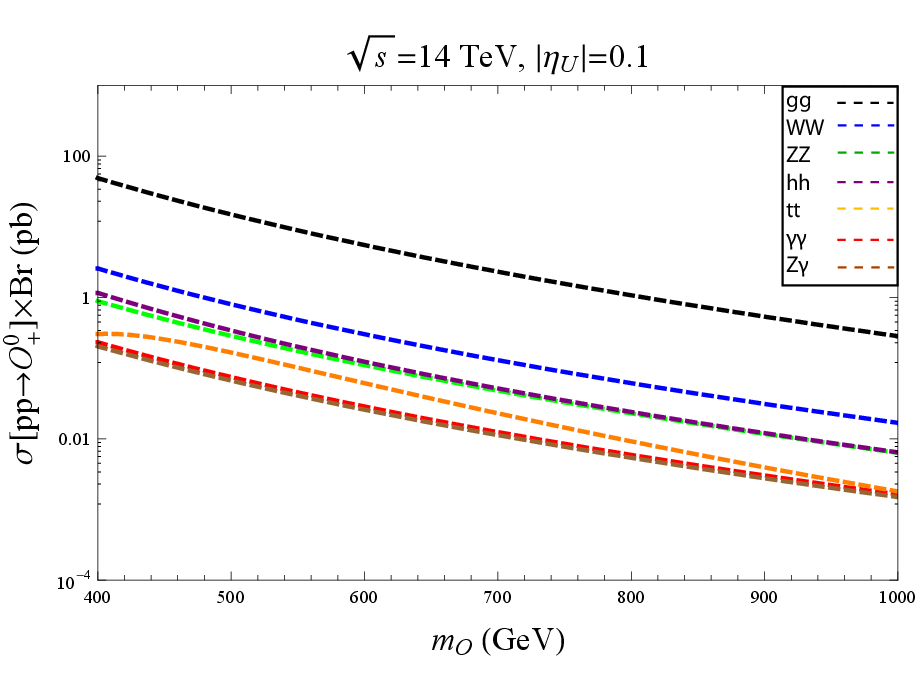} \vspace{-0.5cm}
\caption{LHC cross sections for $pp\to O_+^0  \to X$, where $X$ is a two-body final state.  }
\label{CrossSections1}
\end{figure*}

\begin{figure*}[!t]
\vspace{-1cm}
\includegraphics[width=16 cm]{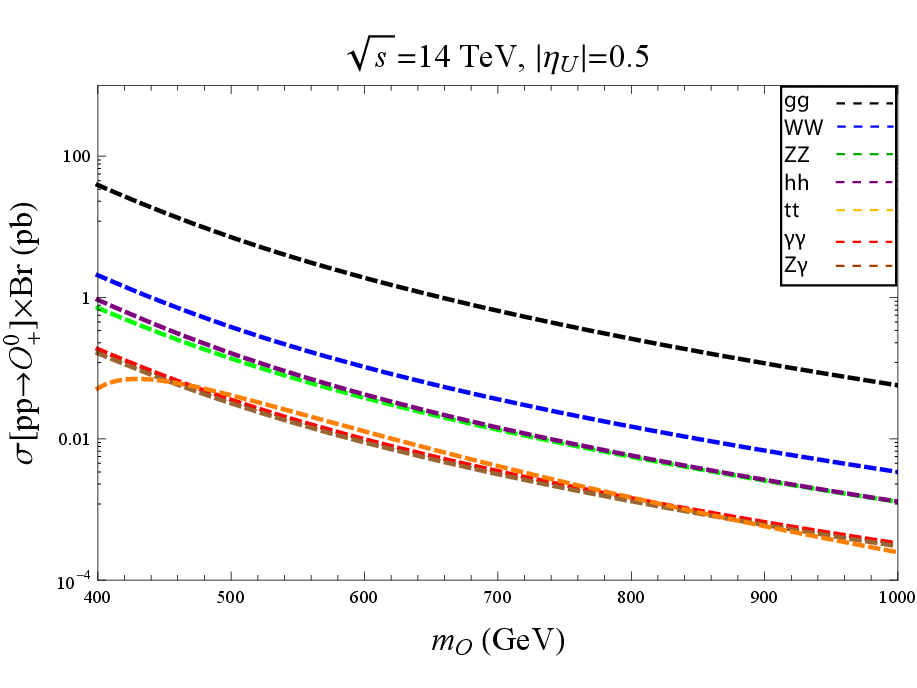} \vspace{-0.5cm}
\caption{LHC cross sections for $pp\to O_+^0  \to X$, where $X$ is a two-body final state.}
\label{CrossSections2}
\end{figure*}

\begin{figure}[!t]
\begin{center}
\vspace{-2cm}
\includegraphics[width=17 cm]{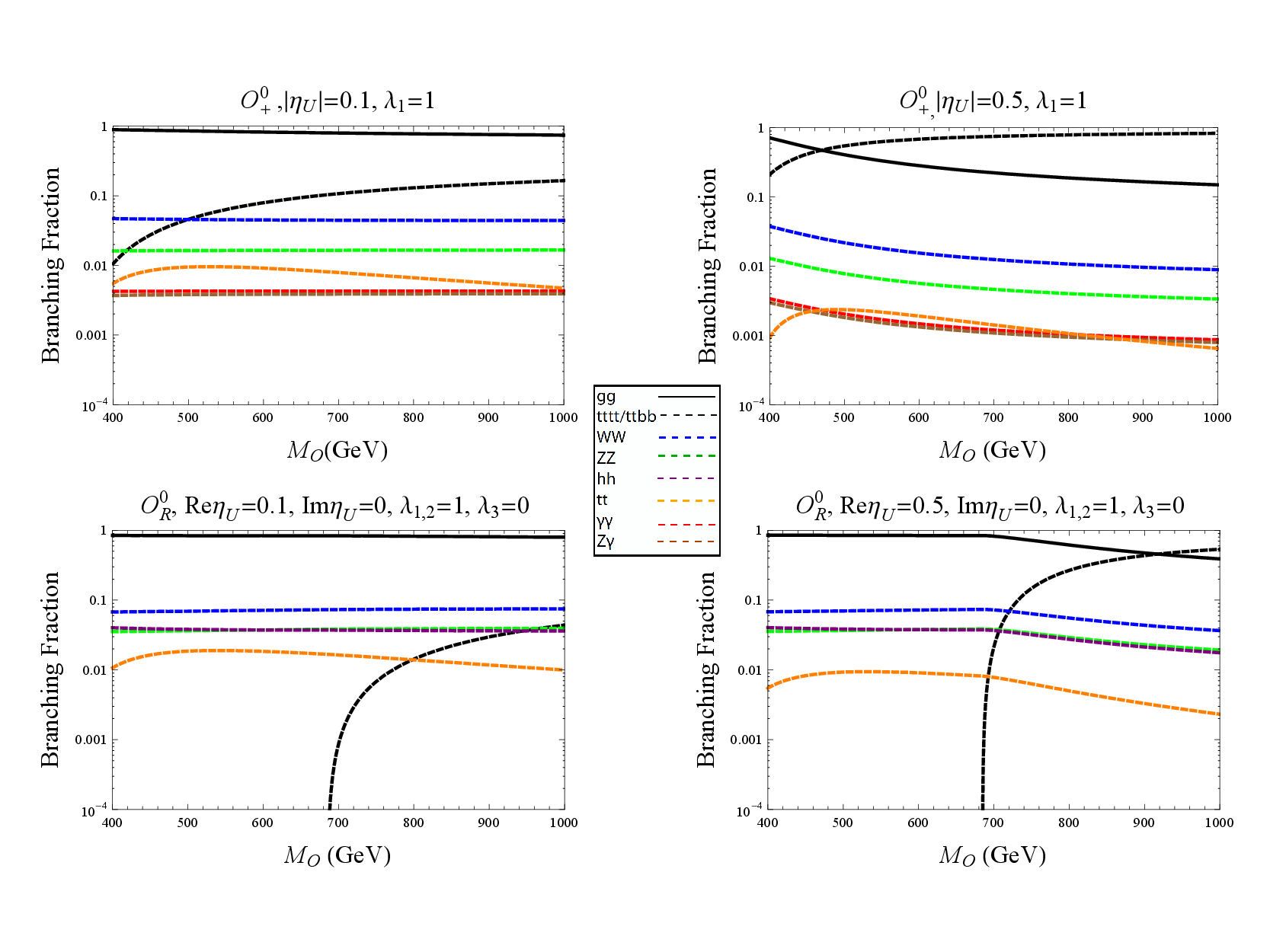} \vspace{-2.3cm}
\caption{Branching fractions for $O_{+,R}^0$ two-body decays for $|\eta_U|=0.1$ and $|\eta_U|=0.5$. \label{Branchingfractions}}
\end{center}
\end{figure}

Explicit formulae for the two-body decay rates are given in the Appendix \ref{appA}. In
Fig.~\ref{CrossSections1} we show the  total cross section  for $O^0_+$ times branching
fractions for the two-body decay modes. In this plot we choose $|\eta_U|=0.1$ and $\lambda_1 = 1.0$. In
Fig.~\ref{CrossSections2} we show the same cross sections  for $|\eta_U|=0.5$ and $\lambda_1 =
1.0$.   The branching fractions for $O^0_+$ and $O^0_R$ as a function of $m_O$ are shown in
Fig.~\ref{Branchingfractions}. In the calculations of $O^0_R$  branching ratios we set
$\lambda_1 = \lambda_2=1.0$, $\lambda_3=0$, Im $\eta_U \equiv \eta_U^I = 0.0$, and Re $\eta_U
\equiv \eta_U^R = 0.1$ and $0.5$. Branching fractions for $O^0_I$ are the same as $O^0_R$ after
replacing $\eta_U^R \leftrightarrow \eta_U^I$ and $\lambda_3\to - \lambda_3$.

\begin{figure}
\begin{center}
\vspace{-1.3cm}
\includegraphics[width=17 cm]{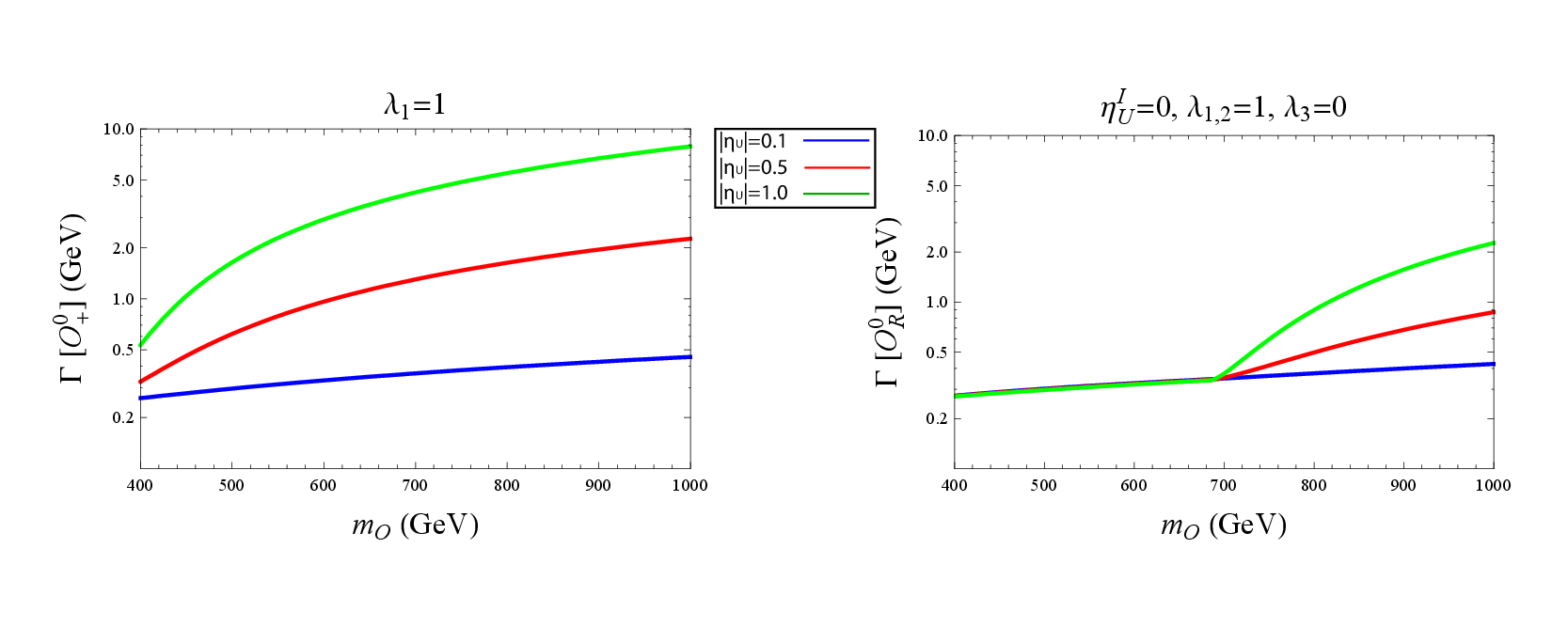} \vspace{-2.3cm}
\caption{Total decay widths of $O_{+,R}^0$ for various values of $|\eta_U|$. \label{DecayRates}}
\end{center}
\end{figure}

For  $|\eta_U|=0.1$, the annihilation decay $O^0_+\to gg$ is the dominant mode for both $O^0_+$ and $O^0_R$. For  $|\eta_U|=0.5$, the decay of $O^0_+$ is
dominated by the non-annihilation decay $O^0_+ \to \bar{t} t \bar{b} b$, while the dominant decay of $O^0_R$ is still $O^0_R \to gg$, because the decay
$O^0_R \to \bar{t} t \bar{t} t $ is phase space suppressed for the masses we are considering, while the decays $O^0_+ \to \bar{t} t \bar{b} b$ and $O^0_+ \to \bar{b} b\bar{b} b$ are suppressed by $m_b^2/m_t^2$  and $m_b^4/m_t^4$, respectively (assuming that $|\eta_U|$ and $|\eta_D|$ are of comparable magnitude).  The total width of
$O^0_+$ and $O^0_R$ as a function of $m_O$ is shown in  Fig.~\ref{DecayRates}, for $|\eta_U|=0.1, 0.5,$ and $1.0$. When the dominant decay channel is $gg$
the widths of the octetonia are less than 1 GeV. When the fall apart decays are dominant, which is the case for $\eta_U =0.5$ or $1.0$ for $O^0_+$,
the  width can be between  1 and 10 GeV depending on $m_{O^0_+}$ and $\eta_U$. For $O^0_R$ the width is always less than 1 GeV
below the $\bar{t}t\bar{t}t$ threshold, and between this threshold and 1000 GeV $\Gamma[O^0_R]\lesssim 2$ GeV even for $|\eta_U|=1.0$.

The next most important decays are to $W^+W^-$ and $Z^0Z^0$. For parameters we have chosen in this paper,
the branching fraction for $O^0_+ \to W^+W^-$ is in the range 
(0.9-4.1)$\times 10^{-2}$ and the branching fraction for $O^0_+ \to Z^0 Z^0$ is in the range (0.3 -1.5)$ \times 10^{-2}$. 
The branching fraction for two Higgs is about the same as $Z^0Z^0$, but this is quite sensitive to the 
Higgs mass (we have chosen 120 GeV in our calculation) and parameters appearing in the potential.
The branching fractions for final states with photons, $\gamma \gamma$ and $\gamma Z^0$, are an order of magnitude smaller
than $W^+W^-$. The branching fraction to $\bar{t}t$ is also very sensitive to model parameters. 
For $|\eta_U|=0.1$ the branching fraction to $\bar{t}t$ is slightly greater than $\gamma \gamma$ and $\gamma Z^0$,
while for  $|\eta_U|=0.5$ the $\bar{t}t$ branching fraction is comparable in size. This unusual result
is due to an accidental cancellation between the two diagrams contributing to $O^0_+ \to \bar{t}t$, see Eq.~(\ref{tt}).
The pattern of decays for $O^0_R$ is similar to $O^0_+$. The main differences are that fall apart decays to $\bar{t}t\bar{t}t$
are suppressed due to kinematics so other branching fractions are slightly larger, and that decays
to $\gamma \gamma$ and $\gamma Z^0$ are absent because the constituents are electrically neutral.

An important question is whether the peaks in the cross section will be visible above Standard Model backgrounds  for these final states. To test this we consider
the process $pp \to O^0_+ \to \gamma \gamma$. We calculate  $d\sigma/d m_{\gamma\gamma} \times {\rm Br}[O^0_+ \to \gamma \gamma]$, including the width of the
$O^0_+$ which  gives this cross section a Breit-Wigner profile. Then we integrate this differential cross section from  $m_{\gamma \gamma}-\Delta/2$ to $m_{\gamma
\gamma} + \Delta/2$ where $\Delta$ is determined by the energy resolution of the measured photon energies. The ATLAS experiment expects to measure  photon
energies with a resolution of \cite{:2008zzm} \bea \frac{\Delta E_\gamma}{E_\gamma} =\sqrt{ \left(\frac{0.1}{E_\gamma/{\rm GeV}}\right)^2 + 0.007^2} \nn \, . \eea
Estimating that $m_{\gamma \gamma} \approx 2 E_\gamma$ and $\Delta m_{\gamma \gamma} \approx \sqrt{2} \Delta E_\gamma$ yields $\Delta m_{\gamma \gamma} \approx
2.8$ GeV for $m_{\gamma \gamma} = 400$ GeV and $\Delta m_{\gamma \gamma} \approx 5.9$ GeV for $m_{\gamma \gamma} = 1000$ GeV. For simplicity we will set $\Delta =
6$ GeV. In Table~\ref{table}, $\sigma_{\rm res}[pp \to O^0_+ \to \gamma \gamma]$ is the total cross section for $pp\to O^0_+ \to \gamma \gamma$ with $m_{\gamma
\gamma}$ within $\pm 3$ GeV of the resonance peak, and  $\sigma_{\rm SM}[pp \to \gamma \gamma]$ is the leading order Standard Model contribution to $pp \to \gamma
\gamma$  in the same kinematic region. 
(By leading order we mean the tree level contribution to $q\bar{q}\to \gamma\gamma$ and the 
one-loop contribution to $gg\to \gamma \gamma$~\cite{Berger:1983yi}.)
We impose the cuts $|\eta_{1,2}|<2.4$, where $\eta_1$ and $\eta_2$ are the rapidities  of the photons in the final state.
We consider the case $|\eta_U| = 0.1$ and $|\eta_U| = 0.5$ and find that the contribution from the octetonium  resonance equals or exceeds the Standard Model
contribution when the octetonium mass is in the 400 to 1000 GeV range. Note that the experimental  background also includes background from events with jets that
fake photons in the detector. 
When $m_{\gamma \gamma}$ is in the range 115 GeV to 140 GeV, this background is (1.25-1.4)$\times\sigma_{\rm SM}[pp\to
\gamma\gamma]$~\cite{Ganjour:2008rz}, but for larger $m_{\gamma \gamma}$ we expect this background to be smaller.
The resonant cross section will still exceed background when 400 GeV $\leq m_{O^0_+} \leq$  1000 GeV for
$|\eta_U|=0.1$. For $|\eta_U|=0.5$ the resonant contribution will exceed the estimated background when 400 GeV $\leq m_{O^0_+} \leq$ 600 GeV,
and possibly for larger $m_{O^0_+}$ depending on the size of the background from  jets faking photons at these high energies. 
These results encourage us to believe that octetonia with masses less than 1 TeV will lead to visible resonances in
$W^+W^-$, $Z^0Z^0$, $\gamma \gamma$, and $\gamma Z^0$  at the LHC. We leave more detailed calculation
of the relevant differential cross sections to future work.

\begin{table*}[t!]
  \begin{tabular}{|c|c|c|c|c|}
\hline
   & \hspace{0.1 in} $400 \,{\rm GeV}$ \hspace{0.1 in} &  \hspace{0.1 in} $600 \,{\rm GeV}$ \hspace{0.1 in} 
    &  \hspace{0.1 in} $800 \,{\rm GeV}$ \hspace{0.1 in}  &  \hspace{0.1 in} $1000 \,{\rm GeV}$ \hspace{0.1 in}   \\
\hline
\hline
\hspace{0.1 in}$\sigma_{\rm res}[pp \to O_+^0 \to \gamma \gamma](\mathrm{fb}), \,|\eta_U|=0.1$\hspace{0.1 in} & 187  & 24  &  5.1  & 1.4 \\
\hline
\hspace{0.1 in}$\sigma_{\rm res}[pp \to O_+^0 \to \gamma \gamma](\mathrm{fb}), \, |\eta_U|=0.5$\hspace{0.1 in} & 149  & 7.7  &  1.1  & 0.23 \\
\hline
 $\sigma_{\rm{SM}}[pp \to \gamma \gamma] (\mathrm{fb})$ & 12.0 & 2.2 &0.64  & 0.23  \\
\hline 
\end{tabular}
\caption{\baselineskip 3.0ex
$\sigma_{\rm res}[pp \to O_+^0 \to \gamma \gamma]$ is the resonant cross section for photons with $|\eta|<2.4$ and $m_{\gamma \gamma}$
within $\pm 3$ GeV of the resonance mass for different values 
of $m_{O^0_+}$, $\sigma_{\rm SM}[pp \to \gamma \gamma]$ is the  Standard Model contribution in the same region.
 }
\label{table}
\end{table*}

To summarize, we have argued that when the Yukawa coupling, $\eta_U$, of color octet scalars  to Standard Model
fermions is less than one, the color octet scalars will live long enough to form color singlet hadrons called octetonia. 
These hadrons will be visible at the LHC as resonances in $W^+W^-$, $Z^0Z^0$, $\gamma \gamma$, and  $\gamma Z^0$
if the masses of the octetonia are between 400 and 1000 GeV. These channels represent new collider signals for searching 
for color octet scalars. If no resonances are observed in these channels at the LHC, lower bounds
on the masses of color octet scalars can be significantly increased. 

\acknowledgments 

This work was supported in part by the U.S. Department of Energy under 
grant numbers DE-FG02-05ER41368 and DE-FG02-05ER41376.
We thank Mark Kruse  and Ahmad Idilbi for helpful discussions
pertaining to this work. We thank Aneesh Manohar and Al Goshaw for comments 
on an earlier version of this manuscript.

\appendix
\section{Partial Decay Rates for Two body decays} \label{appA}
%\section{Appendix}

In this Appendix, we give our results for the partial decay rates for 
the various two-body decays considered in this paper. For decays 
to two gluons or  photons, there are three diagrams involving 
vertices with gauge couplings of the $S^A$ to the vector bosons.
For two-body decays to gluons we find
\bea
\Gamma[O^0_+ \to gg]&=& |\psi(0)|^2 \frac{18  \pi \alpha_s^2(2 m_S)}{m_S^2},  \\
\Gamma[O^0_{\rm R} \to gg] &=& \Gamma[O^0_{\rm I} \to gg] = \frac{1}{2} \Gamma[O^0_+ \to gg].
\eea
For decays to two photons we find,
\bea
\Gamma[O^0_+ \to \gamma \gamma]&=& |\psi(0)|^2 \frac{16 \pi \alpha^2}{m_S^2} \, .
\eea
For decays to $W^+W^-$ and $Z^0 Z^0$, there are diagrams involving 
gauge couplings and also a diagram with a virtual Higgs in the S-channel.
The contribution from these diagrams depends on the parameters
$\lambda_i, i=1,2,$ and $3$, appearing in the formulae for the mass
terms in Eq.~(\ref{masses}). In our numerical calculations we
have chosen the Higgs mass to be $m_h=120$ GeV.
For $W^+W^-$ final states we find 
\bea
&&\Gamma[O_+^0 \to W^+ W^-]=\frac{2|\psi(0)|^2 }{\pi m_S^2} \sqrt{1-x_w}
\Biggl[m_W^4 G_F^2\frac{8-8 x_w +3 x_w^2}{(2-x_w)^2}  \\
&&~~~~~~~~~~~~~~~~~~~~~~~~~~~~~~~~~~~~~~~~~~~~~~~~~
+\frac{3\lambda_1 m_W^2 G_F }{4\sqrt{2}}\frac{x_w}{1-x_h/4}
+ \frac{\lambda_1^2}{128}\frac{4-4x_w +3 x_w^2}{(1-x_h/4)^2}\Biggr], \nn\\
&&\Gamma[O_{R,I}^0 \to W^+ W^-]=\frac{|\psi(0)|^2 }{\pi m_S^2} \sqrt{1-x_w} 
\Biggl[m_W^4 G_F^2\frac{8-8 x_w +3 x_w^2}{(2-x_w)^2} \\
&&~~~~~~~~~~~~~~~~~~~~~~~~~~~~~~~~~~~~~~~~~~~~~~~~~
 +\frac{3\lambda_{R,I} \, m_W^2 G_F}{4\sqrt{2}} \frac{x_w}{1-x_h/4}
+ \frac{\lambda_{R,I}^2}{128}\frac{4-4x_w +3 x_w^2}{(1-x_h/4)^2}\Biggr],
 \nn  
\eea
where $x_{w} =m_{W}^2/m_S^2$, $x_h =m_h^2/m_S^2$, and $\lambda_{R,I} \equiv \lambda_1+\lambda_2 \pm 2\lambda_3$. 
For $Z^0Z^0$ final states we obtain 
\bea
&&\Gamma[O^0_+ \to Z^0 Z^0] =
\frac{|\psi(0)|^2}{\pi m_S^2} \sqrt{1-x_z} 
 \Biggl[  (1-2 s_\theta^2)^4 m_Z^4 G_F^2
\frac{8-8 x_z + 3 x_z^2}{(2-x_z)^2}\\
&&~~~~~~~~~~~~~~~~~~~~~~~~~~~~~~~~~~~~~~~~~~
+\frac{3(1-2 s_\theta^2)^2 m_Z^2 G_F \lambda_1}{4\sqrt{2}}\frac{x_z}{1-x_h/4} 
+ \frac{\lambda_1^2}{128} \frac{4-4 x_z + 3 x_z^2}{(1-x_h/4)^2}
\Biggr], \nn
\\
&&\Gamma[O^0_{R,I} \to Z^0 Z^0] = 
\frac{|\psi(0)|^2}{2\pi m_S^2} \sqrt{1-x_z} \Biggl[ m_Z^4 G_F^2
\frac{8-8 x_z + 3 x_z^2}{(2-x_z)^2} \\
&&~~~~~~~~~~~~~~~~~~~~~~~~~~~~~~~~~~~~~~~~~~~
 +\frac{3m_Z^2 G_F \, \lambda_{R,I} }{4\sqrt{2}}
\frac{x_Z}{1-x_h/4} + \frac{\lambda_{R,I}^2}{128} \frac{4-4 x_z + 3 x_z^2}{(1-x_h/4)^2} \Biggr],\nn  
\eea
where $x_{z} =m_{Z}^2/m_S^2$ and $s_\theta=\sin\theta_W$, where $\theta_W$ is the Weinberg angle. 
Finally, there are decays to $\gamma Z^0$ which do not receive a contribution from the graph with virtual Higgs. 
Our result for these decays is
 \bea
\Gamma[O^0_+ \to \gamma Z^0] = 8\sqrt{2} \,\alpha \,G_F |\psi(0)|^2 
(1-2 s_\theta^2)^2 x_z(1-\frac{x_z}{4})\, .
\eea

The octetonia can also decay to $\bar{t}t$ pairs. There is a tree level
diagram with two Yukawa couplings from Eq.~(\ref{yukawa})
as well as a graph with virtual S-channel Higgs boson.
We find 
\bea\label{tt}
&&\Gamma[O^0_+ \to \bar{t} t]= \frac{3}{32 \pi} \frac{|\psi(0)|^2}{m_S^2}
x_t (1-x_t)^{3/2} 
\left[\frac{\lambda_1}{1-x_h/4} -\frac{4}{N_c}\frac{|\eta_U|^2}{1-x_t}
\frac{m_t^2}{v^2}\right]^2,  \\
&&\Gamma[O^0_{R,I} \to \bar{t} t]= \frac{3}{ \pi} \frac{|\psi(0)|^2}{m_S^2}x_t(1-x_t)^{3/2}\\
&&~~~~~~~~~~~~~~~~~~\times\Biggl[ \left(\frac{2 (\eta_{U}^{R,I})^2}{3}
\frac{m_t^2}{v^2} - \frac{\lambda_{R,I}}{8(1-x_h)}\right)^2 (1-x_t) 
+\frac{4 (\eta_{U}^{R})^2 (\eta_{U}^{I})^2}{9} \frac{m_t^4}{v^4}\Biggr], \nn 
\eea
where $x_t=m_t^2/m_S^2$, $\eta_{U}^{R} = {\rm Re}\,\eta_U$, and $\eta_{U}^{I} = {\rm Im}\,\eta_U$.
Finally, assuming $m_S > m_h$ the octetonia can decay to two Higgs. These proceed
by contact interactions which couple two $S_A$ and two Higgs particles. There is also 
a diagram with S-channel virtual Higgs. 
The decay rates for octetonia to two Higgs are 
\bea
&&\Gamma[O^0_+ \to hh] = \frac{\lambda_1^2}{32 \pi} 
\frac{|\psi(0)|^2}{m_S^2} \frac{(4 + 2 x_h)^2\sqrt{1-x_h}}{(4-x_h)^2},  \\
&&\Gamma[O^0_{R,I} \to hh] = \frac{\lambda_{R,I}^2}{64\pi} 
\frac{|\psi(0)|^2}{m_S^2} \frac{(4 + 2 x_h)^2\sqrt{1-x_h}}{(4-x_h)^2}. 
\eea

%%%%%%%%%%%%%%%%%%%%%%%%%%%%%%%%%%%%%%%%%
%Bibliography

\end{document}